\documentclass[11pt]{article}
\usepackage{color,amsthm,amsmath,amsfonts,xspace}
\usepackage{MnSymbol}
\usepackage{fullpage}
\usepackage{hyperref}
\usepackage[toc,page]{appendix}
\usepackage{authblk}

\newtheorem{Theorem}{Theorem}
\newtheorem{Theorem*}{Theorem}

\newtheorem{Claim*}[Theorem]{Claim}
\newtheorem{Corollary}[Theorem]{Corollary}

\newtheorem{CounterExample*}{$\overline{\hbox{\bf Example}}$}

\newtheorem{Example*}[Theorem]{Example}

\newtheorem{Intuition*}[Theorem]{Intuition}
\newtheorem{Joke*}[Theorem]{Joke}
\newtheorem{Lemma}[Theorem]{Lemma}
\newtheorem{Lemma*}[Theorem]{Lemma}
\newtheorem{Open problem}[Theorem]{Open problem}

\newtheorem{Question*}[Theorem]{Question}
\newtheorem{Remark}[Theorem]{Remark}

\def \bSubexa    {\begin{subexa}}

\newcommand{\ignore}[1]{}

\definecolor{light}{gray}{.75}

\def\ignore#1{}

\newcommand{\bi}{\begin{itemize}}
\newcommand{\ei}{\end{itemize}}

\def\orpro{\mathop{\mathchoice
   {\vee\kern-.49em\raise.7ex\hbox{$\cdot$}\kern.4em}
   {\vee\kern-.45em\raise.63ex\hbox{$\cdot$}\kern.2em}
   {\vee\kern-.4em\raise.3ex\hbox{$\cdot$}\kern.1em}
   {\vee\kern-.35em\raise2.2ex\hbox{$\cdot$}\kern.1em}}\limits}

\def\andpro{\mathop{\mathchoice
 {\wedge\kern-.46em\lower.69ex\hbox{$\cdot$}\kern.3em}
 {\wedge\kern-.46em\lower.58ex\hbox{$\cdot$}\kern.25em}
 {\wedge\kern-.38em\lower.5ex\hbox{$\cdot$}\kern.1em}
 {\wedge\kern-.3em\lower.5ex\hbox{$\cdot$}\kern.1em}}\limits}

\def\simge{\mathrel{%
   \rlap{\raise 0.511ex \hbox{$>$}}{\lower 0.511ex \hbox{$\sim$}}}}

\def\simle{\mathrel{
   \rlap{\raise 0.511ex \hbox{$<$}}{\lower 0.511ex \hbox{$\sim$}}}}

\title{
Universal Variable-to-Fixed Length Lossy Compression at Finite Blocklengths }

\author[1]{Nematollah Iri\thanks{niri1@asu.edu}}

\begin{document}

\maketitle

\begin{abstract}
We consider universal variable-to-fixed length compression of memoryless sources with a fidelity criterion. We design a dictionary codebook over the reproduction alphabet which is used to parse the source stream. Once a source subsequence is within a specified distortion of a dictionary codeword, the index of the codeword is emitted as the reproduced string. Our proposed dictionary consists of coverings of type classes in the boundary of transition from low to high \emph{empirical lossy rate}. We derive the asymptotics of the $\epsilon$-coding rate (up to the third-order term) of our coding scheme for large enough dictionaries. 
\end{abstract}

\section{Introduction}
We consider variable-to-fixed (VF) length compression of an independent and identically distributed (i.i.d.) sequence of random variables $X^{\infty} = X_1 X_2 \cdots$ subject to a fidelity criterion. In VF length coding, a dictionary $\mathcal{D}$ of pre-specified size $M$ is used to parse the source stream $X^{\infty}$. For a $D$-semifaithful VF length code, the source stream $X^{\infty}$ is parsed to subsequences that each are within a given distortion $D$ of a codeword of the dictionary. Note that the dictionary codewords and hence the parsed source subsequences may have variable length. Each variable length parsed subsequence is then mapped to fixed-length ($\log M$ bits) binary representation of the index of its corresponding dictionary codeword.

For a given memoryless source, Tunstall \cite{tunstall} provided an average-case optimal lossless algorithm to maximize average codeword length. A central limit theorem for the Tunstall algorithm's code length has been derived in \cite{drmota}. In most applications, however, statistics of the source are unknown or arduous to estimate, especially at short blocklengths, where there are only limited samples for the inference task. Therefore, we adopt a universal setting where we assume the source distribution $P$ is unknown, while we have the partial knowledge that it belongs to a model class $\mathcal{P}$ --- the collection of memoryless sources over a finite alphabet $\mathcal{X}$ satisfying mild smoothness conditions on the rate-distortion function.

For block source coding in which the source sequence is of fixed length, Shannon \cite{shannon, shannonFid} showed that the rate-distortion function characterizes the best achievable compression performance in the limit of large blocklength. In particular, in the asymptotics of \emph{large blocklength} $n$, sequences $X^n$ drawn from a given i.i.d. source $P$, can be reconstructed within \emph{expected} distortion $D$ by a binary sequence of length greater than $nR(P,D)$, where $R(P,D)$ is the rate-distortion function of the underlying source. In a different variant of lossy source coding, a lossy compression in which the reconstructed sequence is within a distortion $D$ of the source sequence with probability one, is known as the $D$-semifaithful code. Rate-distortion function remains a fundamental limit for semifaithful codes by providing a lower bound on the number of $D$-balls to cover sequences of length $n$. For $D$-semifaithful coding with a known source statistic achievable schemes to the rate distortion function are derived in e.g. \cite{OrnsteinShields, Kieffer, Pursley}. Subsequently, the speed of convergence to the rate-distortion function is studied in \cite{zhangyangwei} wherein an achievable \emph{expected} rate of $R(P,D)+ \frac{\ln n}{n}+ o\left(\frac{\ln n}{n}\right)$ as well as a converse bound of $R(P,D)+\frac{\ln n}{2n}+o\left(\frac{\ln n}{n}\right)$ is derived. For universal block source coding, on the other hand, Yu and Speed \cite{yuspeed} derived an achievable $D$-semifaithful code of redundancy (difference between expected rate and the rate distortion function) $(|\mathcal{X}||\hat{\mathcal{X}}| + |\mathcal{X}|+4) \frac{\ln n}{n} + \mathcal{O}(\frac{1}{n})$.

Modern applications in the 5G and IoT era, however, require compression within a limited latency (delay) and complexity. This requires a more refined metric than redundancy. We evaluate the performance of the VF length code using the $\epsilon$-coding rate --- the minimum rate such that the corresponding overflow probability is less than $\epsilon$. This non-asymptotic figure of merit characterizes the (1-$\epsilon$) percentile of the code length. Therefore, in contrast to all the aforementioned work which optimize the average codelength, non-asymptotic analysis attempts to optimize the cumulative distribution function of the codelength. This refined non-asymptotic analysis consequently provides central-limit theorem type results that characterize the source dispersion --- penalty over the rate-distortion function for compression at a \emph{short blocklength}. For block source coding, the source dispersion has been characterized in \cite{ingberKochman, kostinaVerdu} for non-asymptotic lossy source coding. In this work, we conduct a non-asymptotic analysis for VF length universal coding and derive the dispersion. We further derive the third-order coding rate which is the price of universality for lacking the precise knowledge of the underlying source.

From an algorithmic perspective, the main theme of existing universal VF lenth codes is to define a notion of complexity/information for the source sequences. For lossless compression, such a notion of complexity could be the probability under mixture model \cite{krichevsky, Lawrence, tjalkens, viswer}, or based on the empirical entropy\cite{merhavVFvsFV}. The dictionary is then consists of sequences in the bounaries of transition from low complexity to high complexity. We follow a similar theme  to design the dictionary for the lossy compression. We consider type classes whose empirical lossy rate (to be defined later) is in the transition from low to high rate. We then use the type covering lemma to cover such \emph{transitional} types. Our designed dictionary consists of the covering (reproduction) codewords for transitional types. 

Our goal is to analyze asymptotics of the  $\epsilon$-coding rate as the size of the dictionary increases. We provide an achievable scheme for lossy compression of the class of all finite alphabet memoryless sources $\mathcal{P}$. We provide performance guarantee for VF length $D$-semifaithful compression of finite alphabet memoryless sources using the proposed dictionary. We upper bound the $\epsilon$-coding rate of the proposed VF length code by
\begin{equation*}
	R(P,D) + \sigma(P,D)\sqrt{\frac{R(P,D)}{\log M}}Q^{-1}(\epsilon) + C_{\texttt{Third}}R(P,D)\frac{\log\log M}{\log M} + \mathcal{O}\left(\frac{1}{\log M}\right)
\end{equation*}
where $\sigma^2(P,D)$ is the excess distortion dispersion of the source as defined in \cite{ingberKochman}, $C_\texttt{Third}=\Upsilon+|\mathcal{X}|-1+C_H(1+|\mathcal{X}|)$ in which $\Upsilon$ is a constant provided by the type covering lemma, $C_H$ is a uniform upper bound for the Frobenius norm of the Hessian of $R(P,D)$ and $Q(.)$ is the tail of the standard normal distribution.

Appealing counterpart of this result for the lossless compression of $k$-dimensional parametric model class has been previously derived in \cite{iriVFKosut}:
\begin{equation*}
H+\sigma\sqrt{\frac{H}{\log{M}}}Q^{-1}(\epsilon)+H\frac{k}{2}\frac{\log\log{M}}{\log{M}}+\mathcal{O}\left(\frac{1}{\log{M}}\right)
\end{equation*}
where $H,\sigma^2$ are the entropy and varentropy of the source, respectively.

The remainder of the paper is organized as follows: In Sec. \ref{sec::Notations}, we introduce the notations and formally state the problem. In Sec. \ref{Sec::CodeConstruction}, we present our proposed lossy VF length coding scheme. Technical machinaries required to prove the Main result are provided in Sec. \ref{sec::TechnicalMachinery}. We recall the type covering lemma and state the main result is stated in Sec. \ref{sec::MainThm}. Proof of the main result is provided in Sec. \ref{sec::ProofofMainRes}. We conclude in Sec. \ref{sec::Conclusion}. 

\section{Notations and Problem Statement}
\label{sec::Notations}
 For a set $A$, $|A|$ denotes its cardinality. Let $A^n$ be the $n$-th cartesian product of $A$ and $A^{\infty}=\bigcup_{n=0}^{\infty} A^n$. For two sequences $w^n = w_1,\cdots w_n$ and $z^m=z_1\cdots z_m$, let $w^nz^m$ be the concatenation of the two. Let $\mathcal{P}$ be the collection of \emph{smooth} i.i.d. distributions over the source alphabet $\mathcal{X}=\{1,2,\cdots, |\mathcal{X}|\}$. In the sequel we denote $P\in\mathcal{P}$ as the underlying unknown source model. Let $\mathbb{P}$ and $\mathbb{E}$ denote probability and expectation with respect to the underlying source model $p$. Let $\mathbb{E}_{XY}$ be the expectation with respect to joint distribution of random variables $X$ and $Y$. For a distribution $Q$, let $Q^n$ be the distribution over $n$-length i.i.d. random variables induced by $Q$, i.e. $Q^n(y^n) = \prod_{i=1}^n Q(y_i)$. Denote $P_i = \mathbb{P}(X=i)$ for all $i\in\mathcal{X}$. Let $\mathcal{P}_n$ be the set of all empirical probability mass functions on sequences of length $n$, $n=1,2,\cdots$. For a source sequence $x^n$, we denote its empirical probability mass function as $Q_{x^n}$, i.e. $Q_{x^n}(i) = \frac{n_i}{n}$, where $n_i$ is the number of occurences of the source letter $i$ in $x^n$. Type class of a sequence $x^n$ is the set of all $n$-length sequences of the same empirical probability mass function, i.e. $T_{x^n} = \{z^n: Q_{z^n} = Q_{x^n}\}$. Therefore, all the sequences within the type class $T$ has the same empirical probability mass function which we denote by $Q_T$. $Q_T$ and $Q_{x^n}$ should be distinguished based upon the attribution to a type class $T$ or a source sequence $x^n$. Let $\mathcal{T}_n$ be the set of all type classes over sequences of length $n$, $n=1,2,\cdots$. 
 
 We assume both the source alphabet $\mathcal{X}$ and the reproduction alphabet $\hat{\mathcal{X}}$ are finite. Moreover, without loss of generality we assume $\mathcal{X} = \hat{\mathcal{X}}$. Let $d: \mathcal{X}\times \hat{\mathcal{X}} \rightarrow [0,\infty)$ be a single letter distortion measure. 
 We assume a separable distortion measure. Precisely, the distortion between length-$n$, $n=1,2, \cdots$ source and reproduction sequences $x^n$ and $y^n$ is defined as 
 \begin{equation*}
 	d_n\left(x^n,y^n\right) = \frac{1}{n} \sum_{i=1}^n d(x_i,y_i).	
 \end{equation*}
 
 For block source coding where the source stream $X^n$ to be compressed is in blocks of fixed length $n$, shannon showed that \cite{shannon, shannonFid} minimal expected compression rate within average distortion $D$  is asymptotically bounded below by the rate distortion function of the source defined as 
 
 \begin{equation}
 	R(P,D) := \inf_{\substack{{P_{Y|X}}\\ {\mathbb{E}_{XY}(d(X,Y))\leq D}}} I(X;Y) \label{rateDistIfunc}
 \end{equation}
 where minimization is taken over all pairs of random variables $(X,Y)$ taking values in $\mathcal{X}\times \hat{\mathcal{X}}$ such that $X$ is distributed according to source $P$ and expected distortion between $X$ and $Y$ is less than $D$. 
 
 We depart from the block coding setup by considering $D$-semifaithful VF length coding. A VF length encoder $\phi$ consists of a parsing dictionary $\mathcal{D}$. Elements of the dictionary (codewords) are variable-length sequences over the reproduction alphabet. Once a source subsequence $x^*$ is found within a distortion $D$ of a codeword $y^*\in\mathcal{D}$, the source subsequence is parsed and mapped to the binary representation of its corresponding codeword $y^*$. Let $\ell(x^*)$ be the length of $x^*$. Clearly $\ell(x^*)=\ell(y^*)$ and $d_{\ell(x^*)}(x^*,y^*)\leq D$. Coding continues for the rest of the stream $x_{\ell(x^*)+1}^{\infty}$. We assume a dictionary of size $M$ and hence the binary representation of $y^*$ within the dictionary is $\log M$ bits. 
 
Let $X^*$ be the \emph{random} first parsed subsequence of the source stream $X^{\infty}$. We adopt a one-shot setting and denote 
\begin{math}
\ell^{\phi}(X^{\infty}) := \ell(X^*).	
\end{math}
We gauge the performance of VF length code $\phi$ with a dictionary of size $M$, through the $\epsilon$-coding rate given by
\begin{equation}
R_M(\epsilon,\phi,P) := \min \left\{R: \mathbb{P}\left(\frac{\log M}{\ell^{\phi} (X^{\infty})} \geq R\right)\leq \epsilon\right\} . \label{epsilonCopdingRate}
\end{equation}
 
\textbf{Problem Statement:} We aim to analyze the behavior of $R_M(\epsilon,\phi,p)$ for large enough dictionary size $M$.

\section{Code Construction}
\label{Sec::CodeConstruction}
In this section, we propose our scheme to design the parsing dictionary. We first extract transitional type classes in the boundary of transition from low to high empirical lossy rate. One might think of empirical lossy rate as a measure of the richness of a type class. More precisely, let $\gamma$ be chosen as the lagest positive constant such that the resulting dictionary has at most $M$ codewords; we characterize this $\gamma$ in Subsection \ref{sec::ThresholdSdes}. The type $T$ is a transitional type class if and only if

\begin{equation}
	\label{CodeCosntructuonEq}
	nR\left(Q_T,D\right)\leq\gamma \: \text{and} \:\exists x^n\in T \text{ and } x_{n+1}\in\mathcal{X} \text{ such that } (n+1)R\left(Q_{x^nx_{n+1}},D\right) > \gamma
\end{equation}  
where $R\left(Q_{x^n},D\right)$ is the rate-distortion function for the empirical distribution $Q_{x^n}$ and distortion level $D$ (empirical lossy rate). Similarly define $R\left(Q_T,D\right)$. From lemma \ref{typeCoveringLemma}, let $\mathcal{C}(T)$ be the $D$-covering of type class $T$. We include $\mathcal{C}(T)$ in the dictionary, i.e. 

\begin{equation}
\mathcal{D} = \left\{\mathcal{C}(T): T \text{ is transitional}\right\}. \label{DictionaryEqu}
\end{equation}

Intuitively, sequences with large empirical lossy rate contain more information, and hence our code design compresses largest information into a fixed budget of output bits.

\section{Main Result}
\label{sec::MainThm}
The following type covering lemma is key to design the dictionary codewords.
\begin{Lemma}[Type Covering Lemma]\cite{ingberKochman} \label{typeCoveringLemma}
	Let $Q\in\mathcal{P}_n$, with a corresponding type class $T_Q$. If $\left|\frac{\partial R(Q,D)}{\partial D}\right|$ is bounded in some neighborhood of $Q$, then there exists a codebook $\mathcal{C}(T_Q)$ that completely $D$-covers $T_Q$, where for large enough $n$,
	
	\begin{equation}
		\frac{1}{n} \log{\left|\mathcal{C}(T_Q)\right|}  \leq R(Q,D) + \Upsilon \frac{\log{n}}{n}
	\end{equation}
	where $\Upsilon$ is a constant dependent on $|\mathcal{X}|$ and $|\hat{\mathcal{X}}|$.
\end{Lemma}

We borrow the following notations from \cite{ingberKochman, mahmoodArxiv}: let $\frac{\partial R(P,D)}{\partial P}$ be the gradient of the rate-distortion function $R(P,D)$ whenever the rate-distortion function is differentiable w.r.t. its coordinate. Let $R'(i)$ be the $i$-th component of the gradient, i.e.
\begin{equation*}
	R'(i) := \frac{\partial R(Q,D)}{\partial Q_i} \Bigg|_{Q=P}.
\end{equation*}
Consider $R'(i), i=1,\cdots,|\mathcal{X}|$ as the values taken by a random variable $R'(X)$ \cite{ingberKochman}. Similarly, let $\frac{\partial^2 R(P,D)}{\partial p^2 }$ denote the Hessian of $R(P,D)$, i.e. the $|\mathcal{X}|\times |\mathcal{X}|$ matrix of second-order partial derivatives $\left\{\frac{\partial^2 R(P,D)}{\partial P_i \partial P_j}: 1\leq i,j \leq |\mathcal{X}|\right\}$ when it exists. 

Let $\sigma^2(P,D)$ be the excess distortion dispersion as defined in \cite{ingberKochman}
\begin{equation*}
	\sigma^2 (P,D):= \text{Var} \left(R'(X)\right) = \mathbb{E}\left(R'^2(X)\right) - \mathbb{E}^2\left(R'(X)\right).	
\end{equation*}
The following theorem characterizes the $\epsilon$-coding rate up to the third-order term that is achieved by the proposed dictionary  in Sec. \ref{Sec::CodeConstruction}.

\begin{Theorem}
Let $\mathcal{P}$ be a class of stationary memoryless sources over a finite alphabet $\mathcal{X}$ such that for some $D\in (0,\infty)$, the Frobenius norm $ \Bigg\|\frac{\partial^2 R(P,D)}{\partial P^2 }\Bigg\|_F$ is uniformly bounded over $\mathcal{P}$ by a constant $C_H$. Then, for all $P\in\mathcal{P}$, the VF length code $\phi_{\mathcal{D}}$ using the dictionary $\mathcal{D}$ in (\ref{DictionaryEqu}), achieves
\begin{equation*}
R_M(\epsilon,\phi_{\mathcal{D}}, P) \leq 	R(P,D) + \sigma(P,D)\sqrt{\frac{R(P,D)}{\log M}}Q^{-1}(\epsilon) + C_{\texttt{Third}}R(P,D)\frac{\log\log M}{\log M} + \mathcal{O}\left(\frac{1}{\log M}\right).
\end{equation*}
where $C_{\texttt{Third}} = \Upsilon+|\mathcal{X}|-1+C_H(1+|\mathcal{X}|)$ in which $\Upsilon$ is the constant provided by the type covering lemma (Lemma \ref{typeCoveringLemma}).
\end{Theorem}

\begin{Remark}
The condition on boundedness of the Frobenius norm is a common assumption in the universal lossy compression setup (See e.g. \cite[Theorem 2]{yuspeed}). 
\end{Remark}

\section{Preliminary Results} 
\label{sec::TechnicalMachinery}
For a sequence $x^n\in\mathcal{X}^n$, let $B\left(x^n,D\right)$ be a  distortion-$D$ ball around $x^n$ defined as 
\begin{equation*}
	B(x^n,D) := \{z^n\in\hat{\mathcal{X}}^n: d_n(x^n,z^n)\leq D\}.	
\end{equation*}
Measure of a distortion-$D$ ball is a key quantity in lossy compression and plays the role of the \emph{self-information} in the lossless source coding. 

For distributions $P,Q$, we borrow the following operational lossy rate from \cite{yangZhang,yangKieffer}:
\begin{equation}
	R_0(P,Q,D) := \inf_{U} \left(I(X_1;U) + D\left(P_U \| Q\right)\right)	\label{RPQdbasedonI}
\end{equation}
where $X_1$ is a random variable distributed over $\mathcal{X}$ with distribution $P$, $P_U$ is the distribution of auxilary random variable $U$ over $\hat{\mathcal{X}}$, $D(P_U\| Q)$ is the divergence between $P_U$ and $Q$, and the infimum is taken over all $\hat{\mathcal{X}}$-valued random variables $U$ which are jointly distributed with $X_1$ and satisfy $\mathbb{E}_{XU_1}(d(X_1,U))\leq D$. It turns out that it is easier to analyze $R_0(P,Q,D)$ than $R(P,D)$ \cite{yangZhang}. From (\ref{RPQdbasedonI},\ref{rateDistIfunc}) it is clear that \cite{yangZhang} 
\begin{equation}
	R(Q_T,D) = \inf_{Q} R_0(Q_T,Q,D) \label{operInfDef}.
\end{equation}
Moreover, for any $x^n\in\mathcal{X}^n$ and $Q\in\mathcal{P}$, the following result from \cite{yangZhang}, provides an expression for the measure under $Q$ of a $D$-ball centered at $x^n$:
\begin{equation}
	Q^n\left(B(x^n,d)\right) = \exp\left\{\mathcal{O}(1) - \frac{\ln n}{2} - nR_0(Q_{x^n},Q,d)\right\}. \label{qnequationball}
\end{equation}

\begin{Lemma}
	\label{ProbTaylor}
For an arbitrary $x^{n+1}\in \mathcal{X}^{n+1}$, we have
\begin{equation*}
R(Q_{x^{n+1}},D) \leq R\left(Q_{x^{{n+1}^{-1}}},D\right) + \frac{C_0}{n^2}
\end{equation*}	
where $C_0>0$ is a positive constant and $x^{{n+1}^{-1}}$ is derived from $x^{n+1}$ by dropping the last letter.
\end{Lemma}

\begin{Proof}
For notational convenience, let $x^n = x^{{n+1}^{-1}}$. From Taylor's expansion of $R\left(Q_{x^{n+1}},D\right)$ in the neighborhood of $Q_{x^n}$, we have
\begin{equation}
	R(Q_{x^{n+1}},D) = R(Q_{x^n},D) + \frac{\partial R(Q,D)}{\partial Q}\Bigg|_{Q=Q_{x^n}} \left\|Q_{x^{n+1}}-Q_{x^n}\right\|.
\end{equation}
One can show that there exists constants $\lambda_1,\lambda_2>0$ such that $ \frac{\partial R(Q,D)}{\partial Q}\Bigg|_{Q=Q_{x^n}} \leq \lambda_1$ and $\|Q_{x^nx_{n+1}}-Q_{x^n}\| \leq \frac{\lambda_2}{n^2}$. Therefore, there exists a constant $C_0>0$ such that
\begin{equation}
	R(Q_{x^{n+1}},D) = R(Q_{x^n},D) + \frac{C_0}{n^2}.	
\end{equation}

\end{Proof}

\begin{Corollary}[Codeword Length]
There exists a positive constant $C>0$, such that for any codeword $y^n\in \mathcal{D}$, we have 
\begin{equation}
n \leq C\gamma	.
\end{equation}
\end{Corollary}

\begin{Proof}
Let $x^n$ be the first parsed subsequence for which the corresponding $D$-close codeword is $y^n\in\mathcal{D}$. Let $Q$ be a probability distribution over the reproduction alphabet.
From code construction, $x^n$ belongs to a transitional type class  $T$ and $y^n \in \mathcal{C}(Q_T)$. Again, from the dictionary construction, it must hold that $R(Q_T,D) \leq \gamma$. Let $Q^*$ be the distribution achieving the infimum in (\ref{operInfDef}). 
Since $Q_{x^n}=Q_T$, from (\ref{operInfDef}), we have
\begin{align}
nR\left(Q_T,D\right) &= nR_0\left(Q_T,Q^*,D\right) \nonumber \\
					 &= -\log{Q^*}^n\left(B\left(x^n,D\right)\right)-\frac{\ln n}{2} +\mathcal{O}\left(1\right). \label{expansionRzero}
\end{align}
where (\ref{expansionRzero}) is from (\ref{qnequationball}). Hence, $nR(Q_T,D) \leq \gamma$ implies
\begin{equation}
 -\log{Q^*}^n(B(x^n,D))\leq \gamma +\frac{\ln n}{2} - C_1 \label{upperboundfortheqprob}
 \end{equation}
where $C_1>0$ is a positive constant lower bound for the $\mathcal{O}(1)$ term in (\ref{expansionRzero}).  
%
%
%
%
From the code construction there exist $x^n\in T$ and $x_{n+1}\in\mathcal{X}$, such that $(n+1)R\left(Q_{x^nx_{n+1}},D\right) > \gamma$. Therefore, from Lemma \ref{ProbTaylor},  
\begin{equation}
(n+1)\left(R(Q_T,D)+\frac{C_0}{n^2}\right) > \gamma \label{nandoneequation}
\end{equation}
where $C_0>0$ is the constant in lemma \ref{ProbTaylor}. From (\ref{qnequationball},\ref{nandoneequation}), for a constant $C_2>0$ ( which is a constant upper bound for the $\mathcal{O}(1)$ term in (\ref{qnequationball})) we have 
\begin{equation}
(n+1)\left(-\frac{1}{n}\log{{Q^*}^n\left(B(x^n,D)\right)}-\frac{\ln n}{2n}+ \frac{C_2}{n}\right)>\gamma.
\end{equation}
Therefore
\begin{equation}
	-\log{{Q^*}^n\left(B(x^n,D)\right)} > \frac{n}{n+1}\gamma + \frac{\ln n}{2} - C_2. \label{lowerBoundFroQn}
\end{equation}
Lemma follows from (\ref{upperboundfortheqprob},\ref{lowerBoundFroQn}).

\end{Proof}

%
%
%

We use the following lemma from \cite{mahmoodArxiv} to bound an error term in a Taylor expansion of $R(Q_{T},D)$ around $P$. Similar lemmas are also provided in \cite{ingberKochman, yuspeed}.
\begin{Lemma}\cite{mahmoodArxiv}
	For any $a\geq 0$ such that $a^2\geq 2+2|\mathcal{X}|$, for all $P\in\mathcal{P}$ and all $n=1,2,\cdots$, we have
	\begin{equation}
		\sum_{\substack{Q\in\mathcal{P}_n\\ \|Q-P\|> a\sqrt{\frac{\ln n}{n}}}}{P^n(T_Q)} \leq \frac{e^{|\mathcal{X}|-1}}{n^2}.
	\end{equation}
\end{Lemma}

\section{Proof of Theorem}
\label{sec::ProofofMainRes}
\subsection{Threshold Design}
\label{sec::ThresholdSdes}
Choosing a high threshold value of $\gamma$ in (\ref{CodeCosntructuonEq}), results in compressing more information into a fixed budget of output bits. On the other hand, in order to keep the dictionary size below the pre-specified size $M$, $\gamma$ cannot be too high. In this subsection, we characterize the largest value of $\gamma$ for which the resulting dictionary size is below $M$.

%
Let $N_n$ be the number of dictionary codewords with length $n$. Let
\begin{equation}
	\mathcal{A} =\left\{T\in\mathcal{T}_n: nR\left(Q_T,D\right)\leq \gamma, \exists x^n\in T \text{ and } x_{n+1}\in\mathcal{X} \text{ such that } (n+1)R\left(Q_{x^nx_{n+1}},D\right)>\gamma\right\}.
	\label{definiitionOfA}
\end{equation} 
Motivated by \cite[Eq. 3.12]{merhavVFvsFV}, we upper bound $N_n$ as follows:
\begin{align}
	N_n &= \sum_{T\in\mathcal{A}} |\mathcal{C}(T)| \nonumber \\
	&\leq \sum_{T\in\mathcal{A}}2^{nR\left(Q_T,D\right)+\Upsilon\log n} \label{fromTypeCovering} \\
	&\leq |\mathcal{A}|2^ { \gamma+\Upsilon \log n}
\end{align}
where (\ref{fromTypeCovering}) is from the Type-Covering lemma \ref{typeCoveringLemma}.
We show in Appendix \ref{proofOfSizeA} that $|\mathcal{A}|\leq n^{|\mathcal{X}|-2}$. Therefore, 
\begin{equation}
	N_n \leq n^{|\mathcal{X}|-2} 2^ { \gamma+ \Upsilon \log n}.
\end{equation}

We then upper bound the dictionary size as follows:
\begin{align}
	|\mathcal{D}| &= \sum_{n=0} ^{C\gamma} N_n \nonumber \\
	&\leq 2^\gamma\sum_{n=0}^{C\gamma} n^{\Upsilon+|\mathcal{X}|-2} \nonumber \\
	&\leq C_3 2^\gamma \gamma^{\Upsilon+|\mathcal{X}|-1} \label{plugintoyhit}
\end{align}
where $C_3>0$ is a constant. Finally, from  (\ref{plugintoyhit}), one can show that the following choice of $\gamma$ ensures that the dictionary does not contain more than $M$ codewords:
\begin{equation}
	\gamma  = \log M - (\Upsilon+|\mathcal{X}|-1)\log \log M - C_{\gamma}. \label{gammaValue}
\end{equation}
where $C_{\gamma}>0$ is a constant.


\subsection{Coding Rate Analysis}
In this section, we derive an upper bound for the $\epsilon$-coding rate of the code. Let $X^*$ be the random first parsed source subsequence and $\ell(X^*)$ be its length. For notational convenience, denote $n_R = \frac{\log M}{R}$. Since $M$ is large, without loss of generality, assume $n_R$ is an integer. We assume that for all $P\in\mathcal{P}$, the Frobenius norm of $\frac{\partial^2 R(P,D)}{\partial P^2 }$ is bounded within a neighborhood of $P$, i.e. 
\begin{equation}
	\sup_{Q\in \mathcal{N}(P)} \Bigg\|\frac{\partial^2 R(P,D)}{\partial P^2 }\Big|_{P=Q}\Bigg\|_F \leq C_H \label{HessianMatFnormBound}
\end{equation}
where 
\begin{equation}
	\mathcal{N}(P) := \left\{Q\in P: \|Q-P\|\leq \sqrt{2+2|\mathcal{X}|}\sqrt{\frac{\ln n_R}{n_R}}\right\}.
\end{equation}
Similar assumption has been made in \cite{mahmoodArxiv}. 

We now upper bound the overflow probability as follows:
\begin{align}
	\mathbb{P}\left(\frac{\log M}{\ell(X^*)}>R\right) &= \mathbb{P} \left(\ell(X^*) < n_R\right) \nonumber \\
	&= \mathbb{P}\left(\exists \ell< n_R: \ell R\left(Q_{X^{\ell}},D\right)>\gamma\right) \nonumber \\
	&\leq \mathbb{P}\left(n_R R\left(Q_{X^{n_R}},D\right)>\gamma\right) \nonumber \\
	&= \mathbb{P} \left(R(P,D) + \sum_{i=1}^{|\mathcal{X}|}\left(Q_{X^{n_R}}(i) - p_i\right)R'(i) + \frac{1}{2} (Q_{X^{n_R}}-P) \left(\frac{\partial^2 R(P,D)}{\partial P^2 }\Big|_{P=\tilde{P}}\right) (Q_{X^{n_R}}-P)> \frac{\gamma}{n_R}\right) \nonumber \\
	&\leq \mathbb{P} \left(R(P,D) + \sum_{i=1}^{|\mathcal{X}|}\left(Q_{X^{n_R}}(i) - p_i\right)R'(i) + C_{H}(1+|\mathcal{X}|)\frac{\log n_R}{n_R} > \frac{\gamma}{n_R}\right) \nonumber \\
	&\hspace{1in}+ \mathbb{P}\left(\|Q_{X^{\ell}}-P\|^2 >  (2+2|\mathcal{X}|)\frac{\log n_R}{n_R}\right) \nonumber \\
	&\leq \mathbb{P} \left(R(P,D) + \sum_{i=1}^{|\mathcal{X}|}\left(Q_{X^{n_R}}(i) - P_i\right)R'(i) + C_H(1+|\mathcal{X}|)\frac{\log n_R}{n_R} > \frac{\gamma}{n_R}\right) + \frac{e^{|\mathcal{X}|-1}}{n^2_R} \nonumber \\
	&= \mathbb{P}\left(\frac{\frac{1}{n_R}\sum_{i=1}^{n_R} R'(x_k) - \sum_{i=1}^{|\mathcal{X}|}P_iR'(i)}{\frac{\sigma} {\sqrt{n_R}}} > \frac{\frac{\gamma}{n_R} - C_H(1+|\mathcal{X}|)\frac{\log n_R }{n_R} - R(P,D)}{\frac{\sigma} {\sqrt{n_R}}}\right) + \frac{e^{|\mathcal{X}|-1}}{n_R^2} \nonumber \\
	&\leq Q\left(\frac{\frac{\gamma}{n_R} - C_H(1+|\mathcal{X}|)\frac{\log n_R }{n_R} - R(P,D)}{\frac{\sigma} {\sqrt{n_R}}}\right) + \frac{A}{\sqrt{n_R}} + \frac{e^{|\mathcal{X}|-1}}{n_R^2} \label{equationArrayEnd}
\end{align}

where $\tilde{P} = P + \lambda (Q_{X^{n_R}}-P)$ for some $0\leq \lambda \leq 1$. We have \cite{mahmoodArxiv}
\begin{equation*}
	\left(Q_{X^{n_R}}-P\right) \left(\frac{\partial^2 R(P,D)}{\partial P^2 }\Big|_{P=\tilde{P}}\right) \left(Q_{X^{n_R}}-P\right) \leq C_H \|Q_{X^{n_R}}-P\|^2
	\end{equation*}
%
In Appendix \ref{app:CodeRateDer}, we show that for the rate $R$ specified below, (\ref{equationArrayEnd}) and subsequently the overflow probability falls below $\epsilon$:
\begin{equation*}
R=	R(P,D) + \sigma(P,D)\sqrt{\frac{R(P,D)}{\log M}}Q^{-1}(\epsilon) + \left(\Upsilon+|\mathcal{X}|-1+C_H(1+|\mathcal{X}|)\right)R(P,D)\frac{\log\log M}{\log M} + \mathcal{O}\left(\frac{1}{\log M}\right).
\end{equation*}
Due to the definition of $\epsilon$-coding rate, $R_M(\epsilon,\phi,P)\leq R$. This completes the achievability proof.
\section{Conclusion}
\label{sec::Conclusion}
For universal VF length lossy compression of the class of \emph{all} i.i.d. sources over a finite alphabet, we proposed a dictionary to parse the source stream. For such model class, overflow rate of the proposed coding scheme is derived up to the third-order term in the asymptotics of large dictionary size. Extensions of this work to Markov sources as well as parametric model classes which bear some form of structure are interesting future directions of this work.

{\bibliographystyle{IEEEtran}
\bibliography{reputation}}

\begin{thebibliography}{10}
\providecommand{\url}[1]{#1}
\csname url@samestyle\endcsname
\providecommand{\newblock}{\relax}
\providecommand{\bibinfo}[2]{#2}
\providecommand{\BIBentrySTDinterwordspacing}{\spaceskip=0pt\relax}
\providecommand{\BIBentryALTinterwordstretchfactor}{4}
\providecommand{\BIBentryALTinterwordspacing}{\spaceskip=\fontdimen2\font plus
\BIBentryALTinterwordstretchfactor\fontdimen3\font minus
  \fontdimen4\font\relax}
\providecommand{\BIBforeignlanguage}[2]{{%
\expandafter\ifx\csname l@#1\endcsname\relax
\typeout{** WARNING: IEEEtran.bst: No hyphenation pattern has been}%
\typeout{** loaded for the language `#1'. Using the pattern for}%
\typeout{** the default language instead.}%
\else
\language=\csname l@#1\endcsname
\fi
#2}}
\providecommand{\BIBdecl}{\relax}
\BIBdecl

\bibitem{tunstall}
B.~P. Tunstall, \emph{Synthesis of noiseless compression codes}.\hskip 1em plus
  0.5em minus 0.4em\relax Ph.D. dissert., Georgia Inst. of Technol., Atlanta,
  GA, 1967.

\bibitem{drmota}
M.~Drmota, Y.~A. Reznik, and W.~Szpankowski, ``Tunstall code, khodak
  variations, and random walks,'' \emph{Information Theory, IEEE Transactions
  on}, vol.~56, no.~6, pp. 2928--2937, June 2010.

\bibitem{shannon}
C.~Shannon, ``A mathematical theory of communication,'' \emph{Bell Syst. Tech.
  J.}, vol.~27, pp. 379--423, Oct 1948.

\bibitem{shannonFid}
------, ``coding theorems for a discrete source with a fidelity criterion,''
  \emph{IRE Int. Conv. Rec}, vol.~7, pp. 142--163, Mar 1959.

\bibitem{OrnsteinShields}
D.~S. Ornstein and P.~C. Shields, ``Universal almost sure data compression,''
  \emph{Ann. Prob.}, vol.~18, pp. 441--452, 1990.

\bibitem{Kieffer}
J.~Kieffer, ``A unified approach to weak universal source coding,''
  \emph{Information Theory, IEEE Transactions on}, vol.~24, pp. 674--682, 1978.

\bibitem{Pursley}
K.~M. Mackenthun and M.~B. Pursley, ``Variable-rate universal block source
  coding subject to a fidelity constraint,'' \emph{Information Theory, IEEE
  Transactions on}, vol.~24, pp. 349--360, 1978.

\bibitem{zhangyangwei}
Z.~Zhang, E.~H. Yang, and V.~Wei, ``The redundancy of source coding with a
  fidelity criterion. i. known statistics,'' \emph{Information Theory, IEEE
  Transactions on}, vol.~43, no.~1, pp. 71--91, 1997.

\bibitem{yuspeed}
B.~Yu and T.~P. Speed, ``A rate of convergence result for a universal
  d-semifaithful code,'' \emph{Information Theory, IEEE Transactions on},
  vol.~39, no.~3, pp. 813--820, 1993.

\bibitem{ingberKochman}
A.~Ingber and Y.~Kochman, ``The dispersion of lossy source coding,'' in
  \emph{Proc. Data Compression Conf.}, Mar 2011, pp. 53--62.

\bibitem{kostinaVerdu}
V.~Kostina and S.~Verdu, ``Fixed-length lossy compression in the finite
  blocklength regime,'' \emph{Information Theory, IEEE Transactions on},
  vol.~58, no.~6, pp. 3309--3338, Jun 2012.

\bibitem{krichevsky}
R.~Krichevsky and V.~Trofimov, ``The performance of universal encoding,''
  \emph{Information Theory, IEEE Transactions on}, vol.~27, pp. 199--207, 1981.

\bibitem{Lawrence}
J.~Lawrence, ``A new universal coding scheme for the binary memoryless
  source,'' \emph{Information Theory, IEEE Transactions on}, vol.~23, pp.
  466--472, 1977.

\bibitem{tjalkens}
T.~Tjalkens and F.~Willems, ``A universal variable-to-fixed length source code
  based on lawrence's algorithm,'' \emph{Information Theory, IEEE Transactions
  on}, vol.~38, pp. 247--253, 1992.

\bibitem{viswer}
K.~Visweswariah, S.~R. Kulkarni, and S.~Verdu, ``Universal variable-to-fixed
  length source codes,'' \emph{Information Theory, IEEE Transactions on},
  vol.~47, pp. 1461--1472, 2001.

\bibitem{merhavVFvsFV}
N.~Merhav and D.~L. Neuhoff, ``Variable-to-fixed length codes provide better
  large deviations performance than fixed-to-variable length codes,''
  \emph{IEEE Transactions on Information Theory}, vol.~38, no.~1, pp. 135--140,
  1992.

\bibitem{iriVFKosut}
N.~Iri and O.~Kosut, ``Fundamental limits of universal variable-to-fixed length
  coding of parametric sources,'' in \emph{55th Annual Allerton Conference on
  Communication, Control, and Computing (Allerton)}, June 2017, pp. 31--37.

\bibitem{mahmoodArxiv}
A.~Mahmood and A.~B. Wagner, ``Lossy compression with universal distortion,''
  2021, arXiv:2110.07022.

\bibitem{yangZhang}
E.~Yang and Z.~Zhang, ``On the redundancy of lossy source coding with abstract
  alphabets,'' \emph{Information Theory, IEEE Transactions on}, vol.~45, no.~4,
  pp. 1092--1110, May 1999.

\bibitem{yangKieffer}
E.~Yang and J.~C. Kieffer, ``On the performance of data compression algorithms
  based upon string matching,'' \emph{Information Theory, IEEE Transactions
  on}, vol.~44, pp. 47--65, Jan 1998.

\bibitem{oliver}
O.~Kosut and L.~Sankar, ``Universal fixed-to-variable source coding in the
  finite blocklength regime,'' in \emph{Information Theory Proceedings (ISIT),
  2013 IEEE International Symposium on}, 2013, pp. 649--653.

\bibitem{oliverLalithaJournal}
------, ``Asymptotics and non-asymptotics for universal fixed-to-variable
  source coding,'' \emph{Information Theory, IEEE Transactions on}, vol.~63,
  pp. 3757--3772, 2017.

\end{thebibliography}

\appendices
\section{Proof of $|\mathcal{A}|\leq n ^{|\mathcal{X}|-2}$}
\label{proofOfSizeA}
From the definition \ref{definiitionOfA}, let $x^n\in T$ and $x_{n+1}\in\mathcal{X}$ be such that $(n+1)R(Q_{x^nx_{n+1}},D)>\gamma$. Exploiting Lemma \ref{ProbTaylor} we have $(n+1)R(Q_T,D)+ (n+1)C_0/n^2 > \gamma$. From boundedness of $R(Q_T,D)\leq \log{|\mathcal{X}|}$ we have $nR(Q_T,D) + C_6 > \gamma$ for a constant $C_6>0$. Therefore, we have the following subset relationship
\begin{equation}
\mathcal{A} \subset \left\{T\in\mathcal{T}_n: \frac{\gamma}{n}-\frac{C_6}{n} < R(Q_T,D) \leq \frac{\gamma}{n}\right\}	
\end{equation}
Recalling the boundedness of $ \left|\frac{\partial R(Q,D)}{\partial Q}\right|$, the rest of the proof is similar to \cite{oliver,oliverLalithaJournal}, which we omit.


\section{Achievable $\epsilon$-coding Rate}
\label{app:CodeRateDer}
For notational convenience, let $C_5 = C_H(1+|\mathcal{X}|)$. In order for (\ref{equationArrayEnd}) to be less than or equal to $\epsilon$, it must hold that
\begin{equation}
\gamma -C_5\log{n_R} - n_R R(P,D) = \sigma\sqrt{n_R} Q^{-1}\left(\epsilon - \frac{A}{\sqrt{n_R}}-\frac{2|\mathcal{X}|}{n_R^2}\right).
\end{equation}
By substituting the values of $\gamma$ from (\ref{gammaValue}) and $n_R = \frac{\log M}{R}$, along with the Taylor expansion of $Q^{-1}(.)$, we obtain

\begin{equation}
	R\log M - R(P,D)\log M - R(\Upsilon+|\mathcal{X}|-1+C_5)\log\log M  -RC_{\gamma} +C_5R\log R = R\sigma \sqrt{\frac{\log M}{R}} Q^{-1}(\epsilon) + C_6. \label{firstExpansionj}
\end{equation}
In order to cancel out the highest order $\log M$ term, it must hold that $R = R(P,D)+\delta_1$. Substituting this in (\ref{firstExpansionj}), and denoting $d:= \Upsilon+|\mathcal{X}|-1+C_5$ for convenience, as we obtain
\begin{align}
&\delta_1\log{M}  - d\delta_1\log\log M - dR(P,D)\log\log M  - C_{\gamma}R(P,D) - C_{\gamma}\delta_1 + C_5R\log R \nonumber \\ 
&\hspace{2in}= \sigma\sqrt{R(P,D)+\delta_1}\sqrt{\log M}Q^{-1}(\epsilon) +C_7. \label{secondLargestExp}
\end{align}
Taylor expansion of $\sqrt{R(P,D)+\delta_1}$ around $R(P,D)$ yields
\begin{equation}
\sqrt{R(P,D)+\delta_1} = \sqrt{R(P,D)} + C_8\delta_1. \label{tayRollExo}	
\end{equation}
Replacing (\ref{tayRollExo}) in (\ref{secondLargestExp}), and equating the highest order terms, we obtain

\begin{equation}
\delta_1 = \sigma \sqrt{\frac{R(P,D)}{\log M}}Q^{-1}(\epsilon) + \delta_2. \label{delta1Expansion}
\end{equation}

Finally replacing (\ref{delta1Expansion}) in (\ref{secondLargestExp}) and following the same approach, we obtain 
\begin{equation}
\delta_2 = (\Upsilon+|\mathcal{X}|-1+C_5) R(P,D) \frac{\log\log M}{\log M} +\mathcal{O}\left(\frac{1}{\log M}\right).
\end{equation}
\end{document}